# Electrotunable artificial molecules based on van der Waals heterostructures

Zhuo-Zhi Zhang,[1,2,†] Xiang-Xiang Song,[1,2,†] Gang Luo,[1,2] Guang-Wei Deng,[1,2,*] Vahid Mosallanejad,[1,2] Takashi Taniguchi,[3] Kenji Watanabe,[3] Hai-Ou Li,[1,2] Gang Cao,[1,2] Guang-Can Guo,[1,2] Franco Nori,[4,5] and Guo-Ping Guo[1,2,*]

1. *CAS Key Laboratory of Quantum Information, University of Science and Technology of China, Hefei, Anhui 230026, China.*

2. *Synergetic Innovation Center of Quantum Information and Quantum Physics, University of Science and Technology of China, Hefei, Anhui 230026, China.*

3. *National Institute for Materials Science, Namiki 1-1, Ibaraki 305-0044, Japan*

4. *CEMS, RIKEN, Wako-shi, Saitama 351-0198, Japan*

5. *Department of Physics, University of Michigan, Ann Arbor, MI 48109-1040, USA*

†These authors contributed equally to this work.

*Correspondence to: gwdeng@ustc.edu.cn or gpguo@ustc.edu.cn

**ABSTRACT**

Quantum confinement has made it possible to detect and manipulate single-electron charge and spin states. The recent focus on two-dimensional (2D) materials has attracted significant interests on possible applications to quantum devices, including detecting and manipulating either single-electron charging behavior or spin and valley degrees of freedom. However, the most popular model systems, consisting of tunable double-quantum-dot molecules, are still extremely difficult to realize in these materials. We show that an artificial molecule can be reversibly formed in atomically thin $MoS_2$ sandwiched in hexagonal boron nitride, with each artificial atom controlled separately by electrostatic gating. The extracted values for coupling energies at different regimes indicate a single-electron transport behavior, with the coupling strength between the quantum dots tuned monotonically. Moreover, in the low-density regime, we observe a decrease of the conductance with magnetic field, suggesting the observation of Coulomb blockade weak anti-localization. Our experiments demonstrate for the first time the realization of an artificial quantum-dot molecule in a gated $MoS_2$ van der Waals heterostructure, which could be used to investigate spin-valley physics. The compatibility with large-scale production, gate controllability, electron-hole bipolarity, and new quantum degrees of freedom in the family of 2D materials opens new possibilities for quantum electronics and its applications.

# INTRODUCTION

Two-dimensional (2D) materials are regarded as promising materials for next-generation nanodevices because of their unique band structures and properties (*1-4*). Among the family of 2D materials, transition metal dichalcogenides (TMDCs), represented by $MoS_2$, have attracted significant research interests. The distinct properties of TMDCs, such as appropriate band gap (*5*), high mobility (*6-9*), spin-valley locking (*10, 11*) and large spin-orbit coupling (*12*), make them ideal candidates for electronic (*13, 14*), spintronic (*15-17*), and valleytronic (*18-20*) applications.

It is known that controllable carrier confinement allows detection and manipulation of charge, spin, or valley degrees of freedom electrically (*21-25*), whereas most of the spin-valley experiments in TMDCs were performed by optical techniques (*26*). Using electrically controllable carrier confinement, combined with the diverse nature of 2D materials, rich physics down to the single-particle level could be further investigated. Recently, several studies have demonstrated Coulomb blockade behavior in single-electron transistors (*27*) and electrically-confined single quantum dots (*28-30*) in TMDCs. However, defect-induced impurity traps make it difficult to control the confinement (*31*) to realize a single spin-valley qubit using single quantum dots (*12*). Meanwhile, to achieve electrical manipulations of qubits, the ability to have precise control over multiple quantum dots and their interactions using independent gates is essential (*32, 33*). The realization of a controllable double-quantum-dot molecule in TMDCs still remains a significant challenge because of defects and the complicated gate geometry design.

By integrating a collection of state-of-the-art nanofabrication techniques, we here report the realization of electrotunable confinement in a van der Waals heterostructure based on a few-layer $MoS_2$ sandwiched in h-BN, demonstrating a well-controlled double-quantum-dot system. With independently tuned bottom gates, the coupling strength between the left and right dot can be tuned monotonically, from the weakly to strongly coupled regime, in which *the double-quantum-dot molecule evolves into a single-quantum-dot atom*. This monotonic tunability has not been previously reported in any atomically thin 2D materials; thus, this expands the scope of quantum electronic studies. When tuning the heterostructure into the low-density regime, to study the electrical properties of $MoS_2$, the evolution of the Coulomb peaks shows a suppression of the conductance with increasing magnetic field. This is known as Coulomb blockade weak anti-localization. This observation provides an example for using this well-controlled structure to investigate mesoscopic transport phenomena. It could also be applied to electrically manipulate the spin and valley degrees of freedom for electrons/holes in $MoS_2$ systems and beyond, because the purely ohmic contacts, which currently are obstacles limiting the performance of spintronic/valleytronic devices at low temperature (*34*), are not needed. Our study paves the way for future applications in electrically driven spintronics or valleytronics.

**RESULTS**

Figure 1 shows optical and scanning electron microscopy images with schematics of the samples. MoS$_2$ and h-BN films were exfoliated and transferred onto the local bottom gates [labeled "UM" (upper middle), "LB" (left barrier), "LP" (left plunger), "DM" (down middle), "RP" (right plunger), and "RB" (right barrier) in the inset of Fig. 1B], which are prepatterned on a 100-nm silicon oxide layer, covering the highly-doped silicon that acts as the global back gate (BG). After thermal annealing and contact formation, another h-BN flake was transferred onto the structure. Compared to previous studies (*28, 31*), this structure allows a more aggressive influence of gate tuning through atomically thin h-BN, rather than through atomic layer deposition (ALD)-grown dielectric layer. Short-time light illumination was applied using an infrared light-emitting diode in the low-temperature environment before measurement (*29*). These treatments were used to avoid defect-induced impurity traps as much as possible. All the measurements were performed in a He-3 refrigerator with a base temperature of 230 mK.

First, a dc voltage was applied to the global BG to tune the Fermi energy of the MoS$_2$ device without applying any local bottom gates voltage. In Fig. 2A, the output characteristics (source-drain current, $I_{SD}$, versus source-drain voltage, $V_{SD}$) were measured for different back-gate voltages ($V_{BG}$), showing a typical result for an n-doped device and also indicating that an ohmic contact was formed. The field-effect mobility was estimated to be ~300 cm$^2$/(V·s) from transfer characteristics, as shown in fig. S1. Next, we investigated the pinch-off properties of the local bottom gates by applying an identical gate voltage $V_{\text{local-bottom-gates}}$ to all of the bottom gates simultaneously, using the standard lock-in measurement method. Figure 2B shows the current readout at various negative $V_{\text{local-bottom-gates}}$ and positive $V_{\text{BG}}$ above the turn-on threshold voltage. The conductance of the channel can be monotonically tuned by the gates, down to completely pinched off. By applying different gate voltages ($V_{\text{LB}}$, $V_{\text{RB}}$, $V_{\text{UM}}$, and $V_{\text{DM}}$) independently, the device can be tuned to a double-quantum-dot molecule regime, as shown in Fig. 2C, which is a charge stability diagram (CSD) of the quantum dots measured directly via dc transport (*35*). The formation of the double quantum dots agrees with the potential profiles we calculated using a commercial finite-element analysis simulation software (COMSOL) by solving the Poisson equation based on the designed pattern and voltages applied. In Fig. 2D, the closed contours suggest where the double dot may be located.

After achieving quantum confinement, we investigated the tunability of the coupling strength between the left and right dot. Figure 3 shows the evolution of the CSD, where only the gate voltage $V_{\text{DM}}$ is tuned, whereas all other gate voltages were fixed. By tuning $V_{\text{DM}}$ to more negative values, which leads to the rising of the middle barrier (Fig. 3, C to E), the CSD evolves from an array of parallel lines (Fig. 3B, inset ▲), which corresponds to the Coulomb blockade for a large single quantum dot, to a hexagonal array of points (Fig. 3A), which is expected for two coupled quantum dots in series. Electrons can only resonantly tunnel through the double quantum dot while electric potentials in both dots are aligned with the Fermi levels in the reservoirs, resulting in nonzero currents only at the vertices in the CSD (*36*). The current at other regions of the hexagonal array remains pinched, indicating the suppression of high-order cotunneling events, which is evidence of low tunneling rates between the dots and the reservoirs (*35*). Furthermore, by applying various gate voltages $V_{DM}$ to DM, we measured the fractional peak splitting *f* to quantitively compare the amount of interdot coupling. Here, *f* is defined as $f = 2\, \delta S/\delta P$, where $\delta S$ is the diagonal splitting measured between vertices, which is proportional to the measured barrier conductance, and experimentally determines the total interaction energy due to classical interdot capacitance and quantum tunneling, and $\delta P$ is the distance between vertex pairs, which could be considered as a normalization factor (Fig. 3A) (*37, 38*). Both capacitive coupling and tunnel coupling determine *f*. However, along with opening the inter-dot channel, the interdot capacitance increases more slowly than logarithmically with interdot tunnel coupling (*35*). Thus, the change in the tunnel coupling dominates the change in *f*. As shown in Fig. 3B, by only changing the gate voltage $V_{DM}$, which is applied to the gate DM, the coupling strength between the left and right dot can be tuned monotonically, resulting in electrotuning from a double-dot molecule to a single-dot atom. The experimental results were also consistent with our simulations using COMSOL (fig. S5). This remarkable tunability of the coupling strength is essential to further manipulate electrons inside the quantum dot (*39*), which is absent in previous studies of quantum dots based on $MoS_2$-like TMDCs. Note that monotonic tuning is also difficult to achieve in another counterpart of 2D materials: etched graphene nanostructures, where nonmonotonic behavior is usually observed (*40*). The Supplementary Materials show the tunability of the gate DM over a wider range (fig.

S2), similar results obtained from another sample (fig. S3), and details of the curve fitting.

Using similar transport characteristics to traditional semiconductor double dots, we can extract relevant energy factors in different regimes using the constant-interaction model (*35*). In the strongly-coupled regime, finite-bias measurements were taken via dc transport. As shown in Fig. 3F, clear Coulomb diamonds were observed. The charging energy of the "single large dot" can be extracted from the height of the diamond along the $V_{SD}$ axis (~2.5 meV). Using $E_c = e^2/(8\varepsilon_0\varepsilon_r r)$, where $\varepsilon_r$ is the relative permittivity of MoS$_2$ and $r$ is the radius of the quantum dot (*35*), considering that the relative permittivity of eight-layer MoS$_2$ at low external electric fields can be typically determined (*41, 42*) as 7, we can estimate the radius of the quantum dot to be ~128 nm. This size estimation agrees with the design parameters of the local bottom gates. The two sides of the diamonds represent where the energy level in the dot aligns with the Fermi level in the leads. According to the constant-interaction model (*35*), the slope of the two sides can be calculated as $|e|C_g/C_S$ and $-|e|C_g/(C-C_S)$, respectively. The lever arm $\alpha_{LB}$, which describes the tunability of the gate, can be calculated as $\alpha_{LB} = eC_g/C = 0.17$ eV/V. Here, $C_{S(D)}$ is the capacitance between the quantum dot and the source (drain), $C_g$ is the capacitance between the quantum dot and the gate, and $C$ is the total capacitance: $C = C_S + C_D + C_g$. The calculated lever arm is much larger than that obtained from quantum dots based on traditional GaAs/AlGaAs heterostructures (*39*) and on MoS$_2$-like materials using an ALD-grown dielectric layer (*28, 31*), due to the closer distance between split gates and the flake.

The large lever arm here allows more aggressive influence on the charge carriers, suggesting that a lower gate voltage is needed.

In the weakly-coupled regime, shown in Fig. 3G, the so-called triple points expand into triangular regions when a finite-bias voltage of $V_{SD} = -2$ mV is applied. When a single large dot splits into two separate smaller dots, the tunability of the same gate to the dots evolves, which we can extract from this CSD using the constant-interaction model (*35*) according to

$$\alpha_{LB(RB)}^{L(R)} = e \frac{C_{LB(RB)}^{L(R)}}{C_{L(R)}} = e \frac{V_{SD}}{\delta V_{LB(RB)}} \quad (1)$$

where $C_{LB(RB)}^{L(R)}$ is the capacitance between the gate LB (RB) and the left (right) dot. $C_{L(R)}$ is the total capacitance of the left (right) dot. The lever arms are determined to be $\alpha_{LB}^{L} = 0.23$ eV/V and $\alpha_{RB}^{R} = 0.23$ eV/V, which are also larger than those obtained from traditional GaAs systems (*39*) and etched graphene nanostructures (*40*). According to the dimensions of the honeycomb labeled $\Delta V_{LB}$ in Fig. 3G, the capacitance between LB and the left dot can be extracted as $C_{LB} = e/\Delta V_{LB} = 7.6$ aF. Similarly, the capacitance between RB and the right dot is $C_{RB} = e/\Delta V_{RB} = 7.8$ aF. In addition, we can calculate the charging energy, which is $E_C^L = \alpha_{LB}^L \cdot \Delta V_{LB} = 4.8$ meV for the left dot and $E_C^R = \alpha_{RB}^R \cdot \Delta V_{RB} = 4.7$ meV for the right dot. Further, the estimated dot radius was ~68 nm for both dots, which is consistent with the experimental results that a large quantum-dot atom with a radius of ~ 128 nm evolves into a double-quantum-dot molecule with an individual radius of ~ 68 nm by only tuning the gate voltage $V_{DM}$. The coupling energy between the left and right dots can be extracted from the splitting of the triple points according to $E_C^m = \alpha_{LB}^L \cdot \Delta V_{LB}^m =$ $\alpha_{RB}^{R} \cdot \Delta V_{RB}^{m} = 1.2$ meV. These energy factors extracted from our gate-defined quantum dots are comparable with those obtained from the etched graphene quantum dots (*40*).

To reach the few-electron regime, we tried to decrease the value of $V_{BG}$ to reach a low-density regime. However, the gate controllability here is not as good as that in the high-density regime, which was demonstrated above. The effect of tuning $V_{DM}$ at low density is to only change the tunneling rate, whereas the coupling strength remains almost unaffected (fig. S6), indicating that the tunneling between chains of accidental impurity-defined traps dominates here. As shown in Fig. 4 (D and E), at a relatively low Fermi energy ($E_{F1}$), the intrinsic and fabrication-induced impurities dominate the confining potentials of the transport behavior, which cannot be well controlled by electrostatic gating, whereas at a relatively high Fermi energy ($E_{F2}$), the confining potential is dominated and controlled by electrostatic gating.

We investigated Coulomb blockade peaks under magnetic field in the low-density regime (with $V_{BG}$ = 25 V). Representative sets of Coulomb blockade peaks at $B$ = 0 and 1.5 T (Fig. 4, A and B, respectively) show a suppression of the conductance at high fields. The evolution of the average peak height for over 50 Coulomb blockade peaks, as a function of the magnetic field, is shown in Fig. 4C. We observe a decreasing average peak height with increasing magnetic field.

Different from the weak localization found in few-layer MoS$_2$ open systems (*43*), our results suggest Coulomb blockade anti-localization(*44*) occurring in the MoS$_2$ quantum dot, which is also different from the traditional GaAs quantum dots (*45*). By roughly fitting the data using the Hikami-Larkin-Nagaoka model (*46*):

$$\Delta\sigma = \sigma(B) - \sigma(B=0) = \alpha \frac{e^2}{2\pi^2 \hbar}\left[\psi\left(\frac{1}{2} + \frac{B_\phi}{B}\right) - \ln\left(\frac{B_\phi}{B}\right)\right], \quad (2)$$

where $\psi$ is the digamma function, $e$ is the electronic charge, $\hbar$ is the reduced Plank's constant, $B$ is the magnetic field, and $\alpha$ is an empirical fitting parameter. $B_\phi = \frac{\hbar}{4eL_\phi^2}$ is the phase coherence magnetic field, where $L_\phi$ is the phase coherence length. The fitting parameter $L_\phi$ is determined to be $108 \pm 10$ nm, which is larger than the values obtained from open systems in TMDCs (*43, 47, 48*). In addition, $\alpha$ is estimated to be $-0.107 \pm 0.007$. The observation of anti-localization in the conduction band may possibly be due to the presence of short-range disorder (*49*), such as vacancies in the chalcogen atom layer rather than spin-orbit coupling alone (*44, 50, 51*), whereas the detailed physics behind needs to be further explored in the future.

**DISCUSSION**

In conclusion, by integrating a collection of state-of-the-art nanofabrication techniques developed for van der Waals materials, we have demonstrated a gate-controlled artificial molecule system, consisting of two coupled quantum dots, in a few-layer $MoS_2$-based heterostructure. The interdot coupling strength can be tuned monotonically via the control gate. With a purely capacitively coupled dots model, the relevant energy factors could be extracted from our results. The estimated dot size agrees well with the gate design parameters. By analyzing the magnetoconductance evolution in the low-density regime, Coulomb blockade weak anti-localization was observed.

Previous quantum dots in 2D materials were widely studied in graphene (*4*), usually through the plasma etching methods (*52*), which introduces impurities and defects at the etched edge, leading to a limited controllability (*53*) and an enhanced noise level (*54*), thus limiting the performance of the quantum dot. Semiconducting $MoS_2$ makes it possible to achieve quantum dots via an electric field. Interesting physics could be studied if we replace the contacts using metallic, superconducting, ferromagnetic, antiferromagnetic, and ferroelectric 2D materials due to their diversity. Coupling of nearby quantum dots in 2D materials could be achieved not only laterally but also vertically (*55*) with extremely close distance, which may be applied in future integrations. Regardless of lattice mismatch in bulk systems and etching-induced limitation in graphene devices, our demonstration, in which semiconducting and insulating atomically thin materials were hybridized, presents a possible platform for the electrical detection of spin-valley physics and manipulation of various quantum degrees of freedom in the atomic flatland, which were mostly realized by optical methods (*26*). In addition, the large spin-orbit coupling in $MoS_2$ revealed in Coulomb blockade weak anti-localization could be further used to investigate and electrically , rather than magnetically, manipulate the spin/valley degrees of freedom.

## MATERIALS AND METHODS

### Device fabrication

At the beginning of the fabrication process, the bottom gates were prepatterned using standard electron beam lithography (EBL) on a 100-nm-thick silicon oxide layer, which covered the highly-doped silicon that acted as the global BG, as shown in Fig. 1A. Different from traditional semiconductor heterostructures, local bottom gates formed by 5-nm-thick palladium were used, instead of surface gates on top of the dielectric layer, to avoid direct electron beam exposure of the quantum-dot area. $MoS_2$ (bulk from SPI Supplies) and h-BN thin flakes were mechanically exfoliated onto 285-nm-thick layers of silicon oxide to achieve a high contrast, allowing the flakes to be distinguished and selected under an optical microscope (56). $MoS_2$ and h-BN thin flakes with proper thicknesses were subsequently picked up and transferred onto the prepatterned bottom gates using a thin film of polycarbonate (57). After transfer, the whole structure was annealed in an $Ar/H_2$ mixture at 250 °C for 45 min to remove any residual material and bubbles from the transfer process. Then, another EBL step was performed followed by electron beam evaporation of titanium (5 nm thickness) and gold (90 nm thickness) to form the metal contacts to $MoS_2$; these are labeled by "S" (for source) and "D" (for drain) in the inset of Fig. 1B. After the lift-off procedure, another h-BN flake was transferred onto the structure, to protect the upper interface of the $MoS_2$ flake using the same transfer method as described above.

### Measurement setup

All the measurements were performed in a He-3 refrigerator with a base temperature of 230 mK. Both dc tranport and standard lock-in technique were applied in the experiments. For the pinch-off properties presented in Fig. 2B, a standard lock-in technique was applied using an SR830 lock-in amplifier. For all the other transport results, we performed dc transport measurements using an SR570 preamplifier and a Keithley 2015-P multimeter.

**SUPPLEMENTARY MATERIALS**

Supplementary material for this article is available at XXXXX

Supplementary Text

fig. S1. Source-drain current $I_{SD}$ versus the global back-gate voltage, $V_{BG}$.

fig. S2. Tunability of the gate DM over a wider range at $V_{BG} = 30$ V.

fig. S3. Tunability of the gate DM in another similar sample.

fig. S4. COMSOL simulation of the interdot barrier.

fig. S5. COMSOL simulation on the potential well distribution for different values of $V_{DM}$.

fig. S6. Gate controllability in the low-density regime.

References (*58*)

**ACKNOWLEDGEMENTS**

We thank Q. Ma from P. Jarillo-Herrero's group and K. Wang from P. Kim's group for fruitful discussions on the transfer method. We also thank Prof. Z. Han and Prof. S. -L. Li for comments on the manuscript. This work was supported by the National Key Research and Development Program of China (grant no. 2016YFA0301700), the National Natural Science Foundation of China (grant nos. 11625419, 61674132, 11674300, 11575172, and 91421303), the Strategic Priority Research Program of the Chinese Academy of Sciences (grant no. XDB01030000), and the Fundamental Research Fund for the Central Universities. F.N. was partially supported by the: Japan Society for the Promotion of Science (JSPS) (KAKENHI), CREST grant no. JPMJCR1676, and the Sir John Templeton Foundation. The growth of the h-BN crystals was supported by the Elemental Strategy Initiative and conducted by MEXT, Japan and JSPS KAKENHI (grant nos. JP26248061, JP15K21722, and JP25106006). This work was partially carried out at the University of Science and Technology of China Center for Micro and Nanoscale Research and Fabrication. **Author contributions:** Z.-Z.Z., G.L., and H.-O.L. fabricated the samples. Z.-Z.Z., X.-X.S., and G.-W.D. performed the measurements. V.M. ran the simulations. Z.-Z.Z., X.-X.S., G.C., and G.-P.G. analyzed the data. K.W. and T.T. grew the single-crystal h-BN. Z.-Z.Z., X.-X.S., and G.-W.D. prepared the manuscript. G.-P.G., F.N., and G.-C.G. advised on experiments and data analysis and contributed to the manuscript writing. **Competing interests:** The authors declare that they have no competing interests. **Data and materials availability:** All data needed to evaluate the conclusions in the paper are present in the paper and/or the Supplementary Materials. Additional data related to this paper may be requested from the authors.

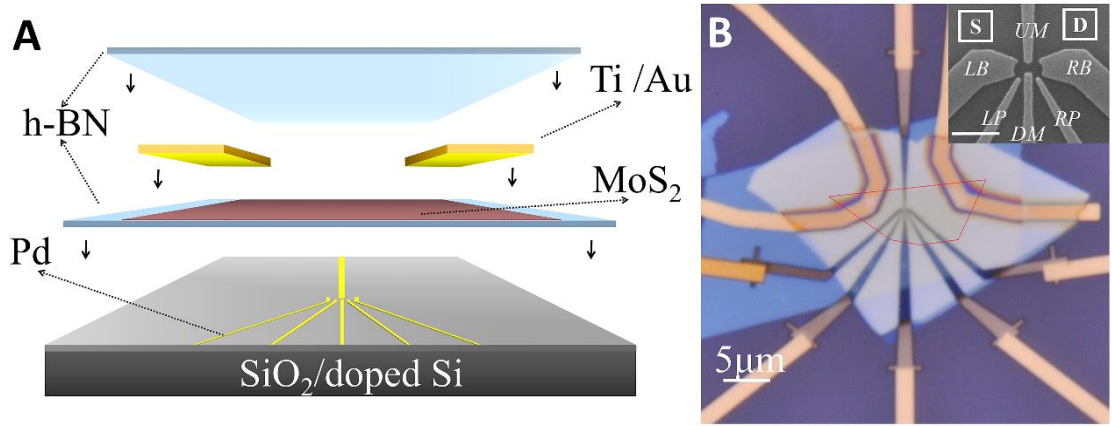

**Fig. 1. Realization of an atomically thin MoS₂ quantum-dot molecule.** (**A**) Schematic diagram of the sample structure. MoS$_2$ flake (determined by atomic force microscopy to be eight layers thick) and source-drain metal [Ti (5 nm)/Au (90 nm)] were sandwiched using two h-BN flakes. (**B**) Optical microscopy image of a typical device with a scale bar of 5 μm. The area enclosed in the red dashed lines indicates the location of the sandwiched MoS$_2$ flake. The inset with a scale bar of 500 nm shows a scanning electron microscopy image of the bottom gate structure taken before the stacks were transferred onto it. The bottom gates are formed by 5-nm-thick Pd. All local bottom gates were applied with negative dc biases to confine the quantum dot, whereas the global BG was applied with a positive dc bias to raise the charge density of MoS$_2$.

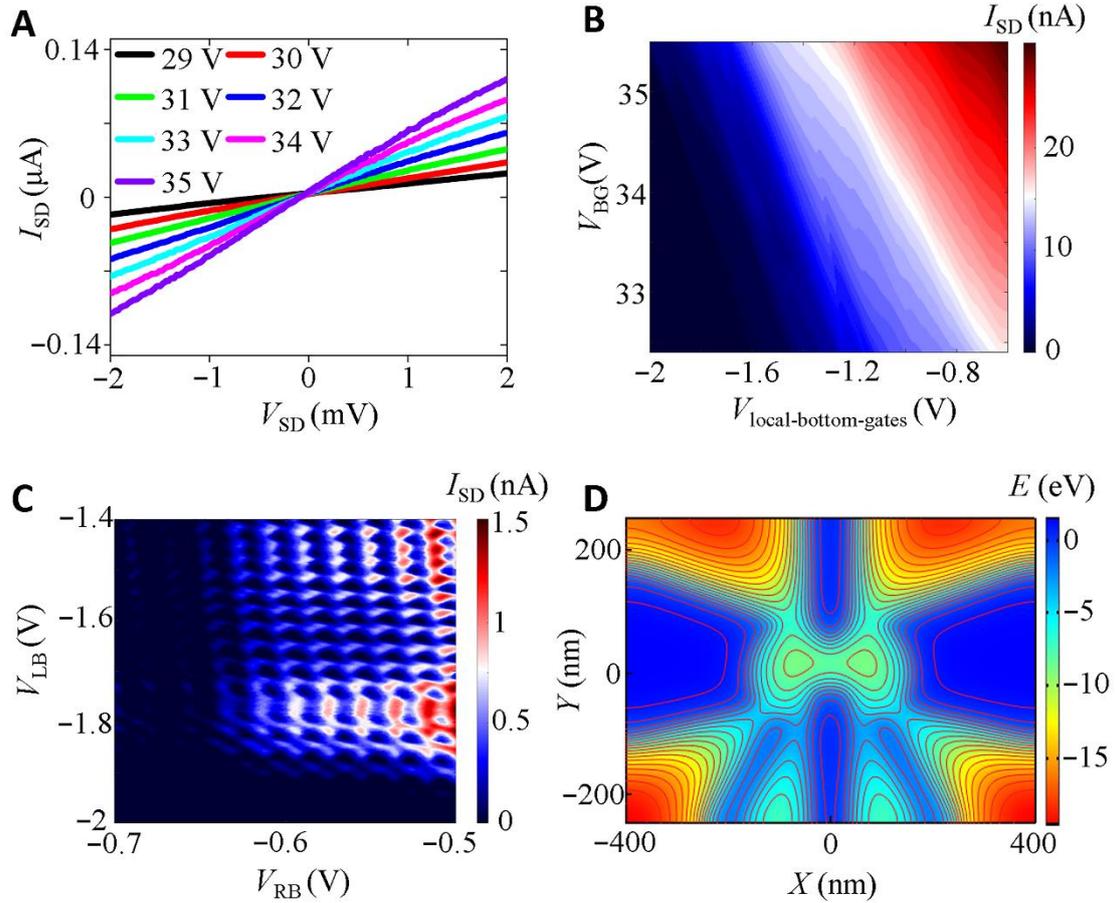

**Fig. 2. Electronic transport characteristics of the MoS$_2$ double quantum dot.** (**A**) Source-drain current, $I_{SD}$, versus source-drain voltage, $V_{SD}$, for various $V_{BG}$ while all local bottom gates are grounded. This shows that, in such a $V_{BG}$ range, an ohmic contact has been formed. (**B**) At zero dc bias and 1 mV ac excitation, a standard lock-in measurement is applied at the source and drain for different $V_{BG}$ and with a uniformly applied local bottom gates voltage ($V_{local\text{-}bottom\text{-}gates}$), indicating a monotonic tuning of the conductance of the channels and the existence of quantum plateaus. (**C**) dc through the double-quantum-dot structure versus $V_{LB}$ and $V_{RB}$, which are applied to the gates LB and RB, for $V_{BG} = 30$ V, $V_{LP} = V_{RP} = 0$ V, $V_{UM} = -1.8$ V, and $V_{DM} = -1$ V and bias voltage at $V_{SD} = 3$ mV. The honeycomb-shaped arrays correspond to the typical transport characteristics of a double quantum dot. (**D**) COMSOL simulation of the potential profile in the MoS$_2$ layer for the gate pattern of the device shown in the inset of Fig. 1B. Here, $V_{BG} = 30$ V, $V_{LP} = V_{RP} = 0$ V, $V_{LB} = V_{RB} = -1.5$ V, $V_{UM} = -1.8$ V, and $V_{DM} = -1$ V. The closed contours indicate where the quantum dots may be located.

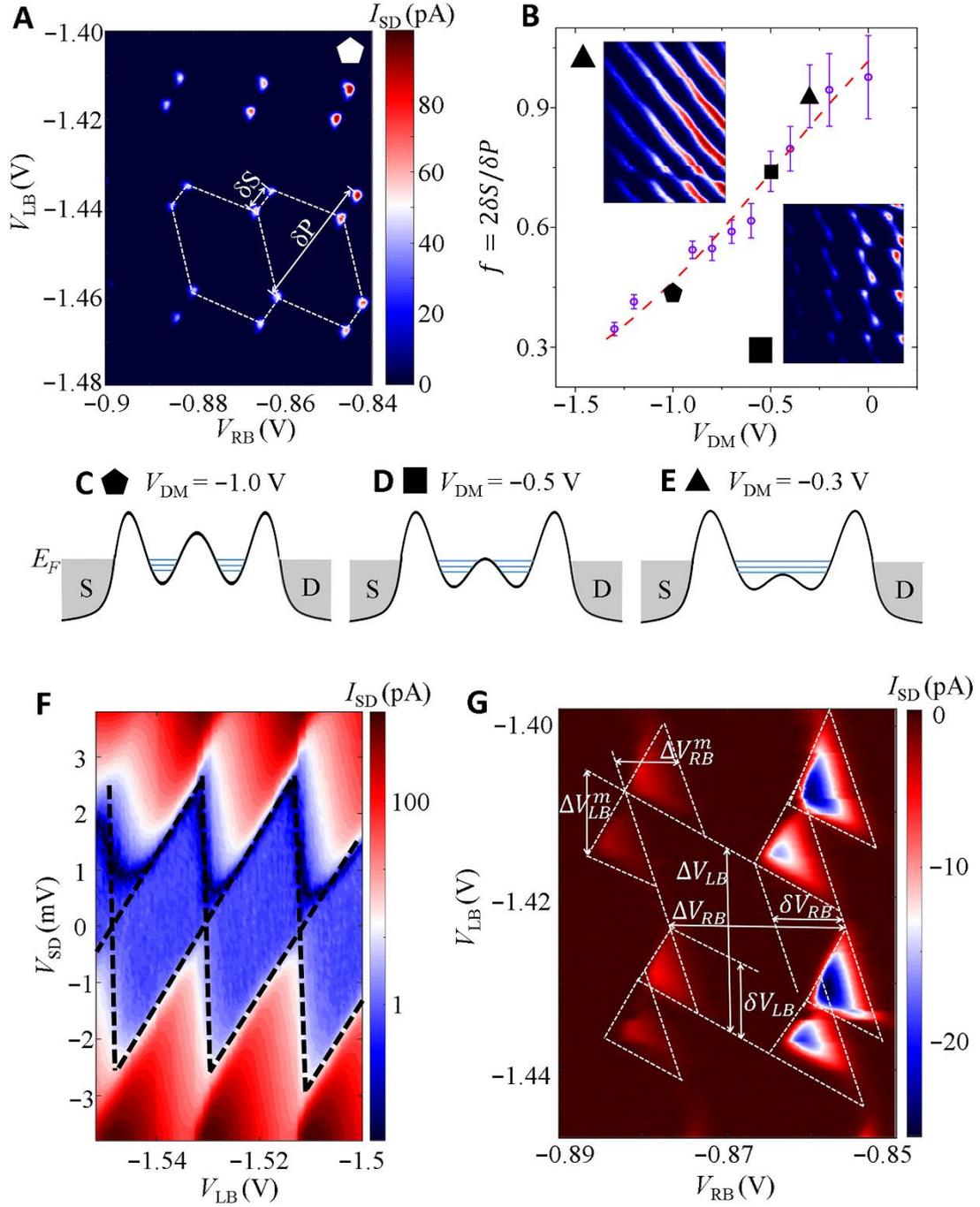

**Fig. 3. Evolution from double to single MoS$_2$ quantum dot by electrostatic gating.**

(**A**) Current through the double quantum dots versus $V_{LB}$ and $V_{RB}$ applied to the gates LB and RB for $V_{BG}$ = 30 V, $V_{LP}$ = $V_{RP}$ = 0 V, and $V_{UM}$ = −2.1 V and bias voltage at $V_{SD}$ = 100 μV and $V_{DM}$ = −1 V; the data demonstrate that the double quantum dot works in the weakly coupled regime. $\delta S$ is the diagonal splitting measured between vertices, and $\delta P$ is the distance between vertex pairs. (**B**) $f = 2\,\delta S/\delta P$ as a function of $V_{DM}$, which shows a decrease of $f$ when the gate voltage $V_{DM}$ applied to gate DM becomes more negative, demonstrating a monotonic change of $f$ by tuning $V_{DM}$. ▲ and ■ indicate CSD at different values of $V_{DM}$, whereas other parameters are kept the same as in (A). Here, the violet circular dots stand for the data points, and the dashed curve is the fitting curve described in the Supplementary Materials. (**C to E**) Energy landscape of the double-quantum-dot system varying from the weakly coupled to the strongly coupled regime, corresponding to ⬟, ■, and ▲, respectively. (**F**) Finite-bias measurements of the double quantum dot in the strongly coupled regime ($V_{BG} = 30$ V, $V_{LP} = V_{RP} = 0$ V, $V_{UM} = -2.1$ V, $V_{RB} = -1$ V, and $V_{DM} = -0.2$ V), where the device behaves as a single large quantum dot. (**G**) Charge-stability diagram at $V_{BG} = 30$ V, $V_{LP} = V_{RP} = 0$ V, $V_{UM} = -2.1$ V, and $V_{DM} = -1.2$ V, as well as for a bias voltage at $V_{SD} = -2$ mV. The triple points expand into triangles; there the lever arm between the gates and the dots and the charging energy of both dots can be extracted as indicated in the figure.

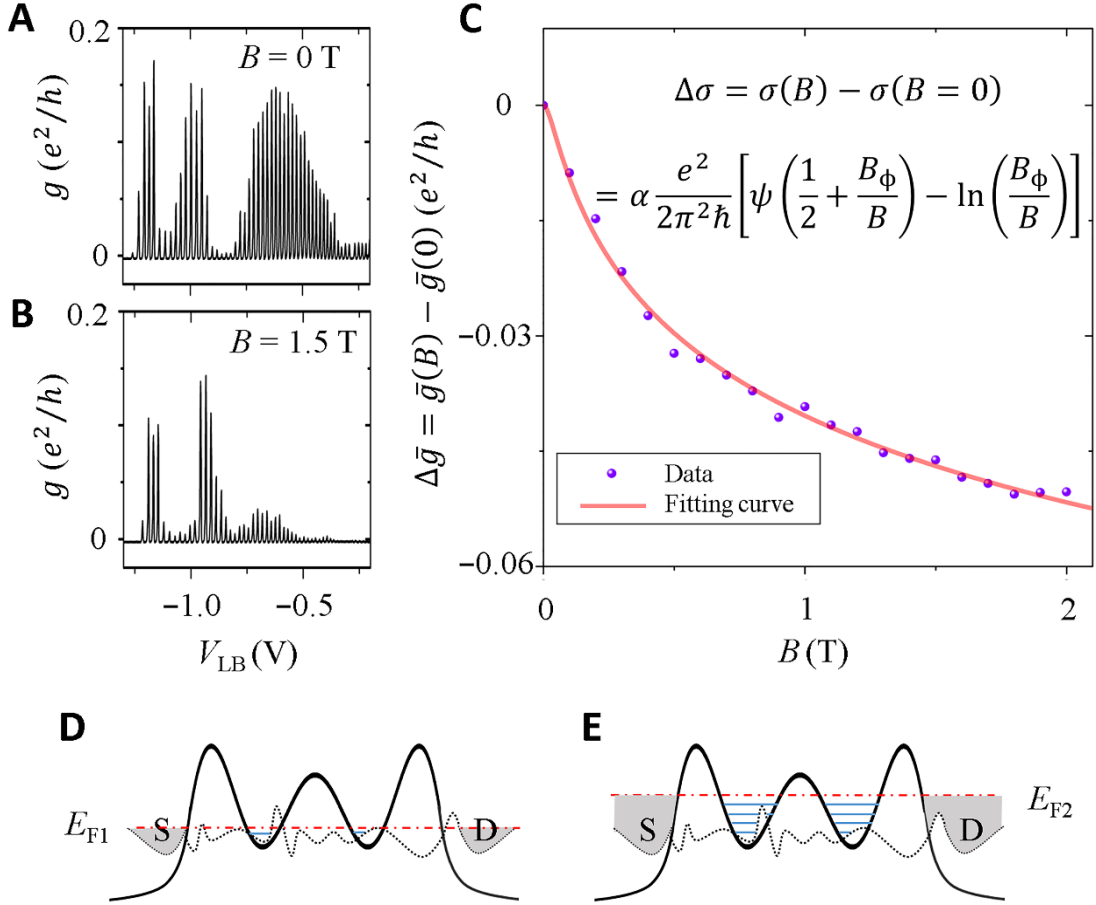

**Fig. 4. Electronic and magnetic transport in the low-density regime.** Representative sets of Coulomb blockade peaks at (**A**) $B = 0$ T and (**B**) $B = 1.5$ T are measured by dc readout with $V_{BG} = 25$ V, $V_{LP} = V_{RP} = 0$ V, $V_{RB} = -0.8$ V, $V_{UM} = -1.5$ V, and $V_{DM} = -1$ V and a bias voltage at 1 mV. (**C**) Evolution of the average peak height for over 50 Coulomb blockade peaks in (A) and (B) versus magnetic field. (**D** and **E**) Energy landscapes of the system varying from the low-density regime to the high-density regime in the presence of a disordered potential, respectively.

Supplementary Materials

*for*

**Electro-tunable artificial molecules**

**based on van der Waals heterostructures**

1. **Source-drain current $I_{SD}$ versus the global back-gate voltage $V_{BG}$**

2. **Tunability of the gate DM over a wider range**

3. **Rough fitting of the fractional peak splitting $f$ versus the gate voltage $V_{DM}$**

4. **COMSOL simulation for different values of $V_{DM}$**

5. **Gate controllability in the low-density regime**

## 1. Source-Drain current $I_{SD}$ versus the global Back-Gate voltage $V_{BG}$

A DC voltage was applied to the global back gate to tune the Fermi energy of the MoS$_2$ device without any local bottom gates voltage applied. The source-drain current ($I_{SD}$) was measured when sweeping the back-gate voltage ($V_{BG}$) at a fixed source-drain voltage ($V_{SD}$) of 5 mV. As shown in fig. S1, the device shows a typical n-type semiconductor behavior. The demonstration of the double quantum dot is presented above the turn-on threshold voltage, with an estimated field-effect mobility of ~ 300 cm$^2$/(V • s). The estimated charge density at $V_{BG}$ = 25 V and 30 V is $2.77 \times 10^{11}$ cm$^{-2}$ and $1.20 \times 10^{12}$ cm$^{-2}$, respectively.

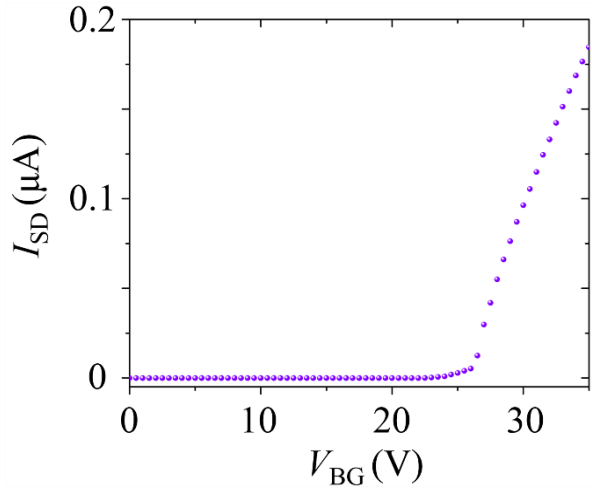

**fig. S1.** Without any local bottom gates voltage applied, the source-drain current $I_{SD}$ was measured when sweeping the global back-gate voltage $V_{BG}$. The $I_{SD}$ versus $V_{BG}$ curve shows a typical n-type semiconductor behavior.

## 2. Tunability of the gate DM over a wider range

When tuning $V_{DM}$ to a negative range, varying from −0.2 V to −1.2 V, while all other gate voltages remain fixed, one large quantum-dot atom evolves into a double-quantum-dot molecule. Figure S2 shows a typical area of the charge stability diagram, which is much larger than that shown in Fig. 3. Figure S3 shows a similar phenomenon observed in another sample.

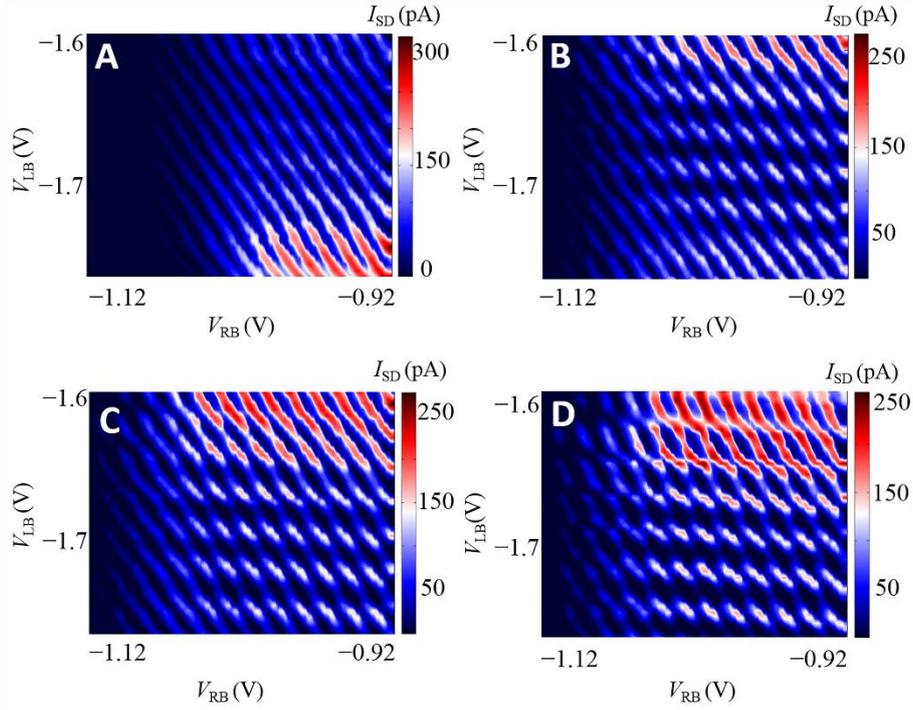

**fig. S2.** Current through the double quantum dot versus $V_{LB}$ and $V_{RB}$ applied to the gates LB and RB for $V_{BG}$ = 30 V, $V_{LP}$ = $V_{RP}$ = 0 V, $V_{UM}$ = −2.1 V, bias voltage at $V_{SD}$ = 100 μV and $V_{DM}$ = −0.2 V, −1 V, −1.1 V and −1.2 V for **A** to **D**, respectively.

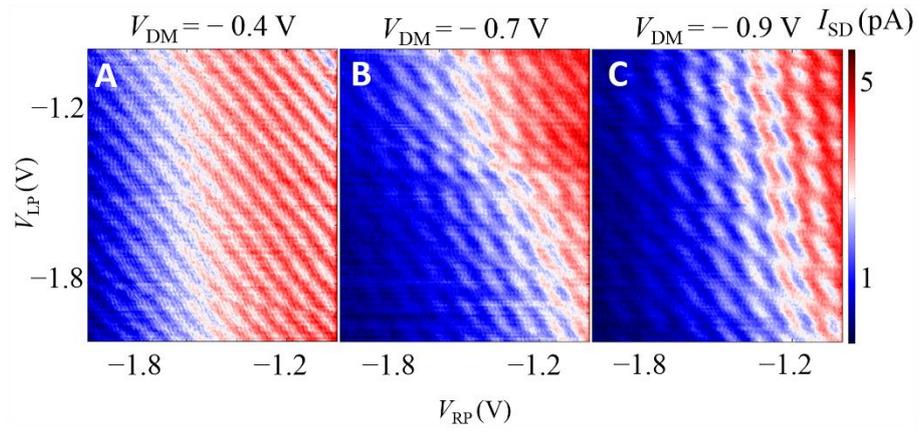

**fig. S3.** Current through the double quantum dot versus $V_{LP}$ and $V_{RP}$ applied to the gates LP and RP, for $V_{BG}$ = 40 V, $V_{LB}$ = $V_{RB}$ = −0.64 V, $V_{UM}$ = −1.1 V, bias voltage at $V_{SD}$ = 100 μV and $V_{DM}$ = −0.4 V, −0.7 V and −0.9 V for **A** to **C**, respectively. This data was measured from another sample with same structure.

### 3. Rough fitting of the fractional peak splitting $f$ versus the gate voltage $V_{DM}$

Consider the inter-dot potential barrier in a parabolic form, $U(x) = -\frac{1}{2}kx^2$, the inter-dot coupling strength which corresponds to the transmission coefficient $D$ can be determined (58) as $D = 1/(1 + e^{-2\pi\epsilon})$, where $\epsilon = (E/\hbar)\sqrt{m/k}$. We can roughly fit the fractional peak splitting $f$ versus the gate voltage $V_{DM}$ with a $1/(1 + e^{-kx})$ lineshape. It is worth noticing that in the strongly-coupled regime, the inter-dot potential barrier does not fit the quasi-classical model because of the high value of the transmission coefficient. So the fit is not suitable when the value of $V_{DM}$ is near 0.

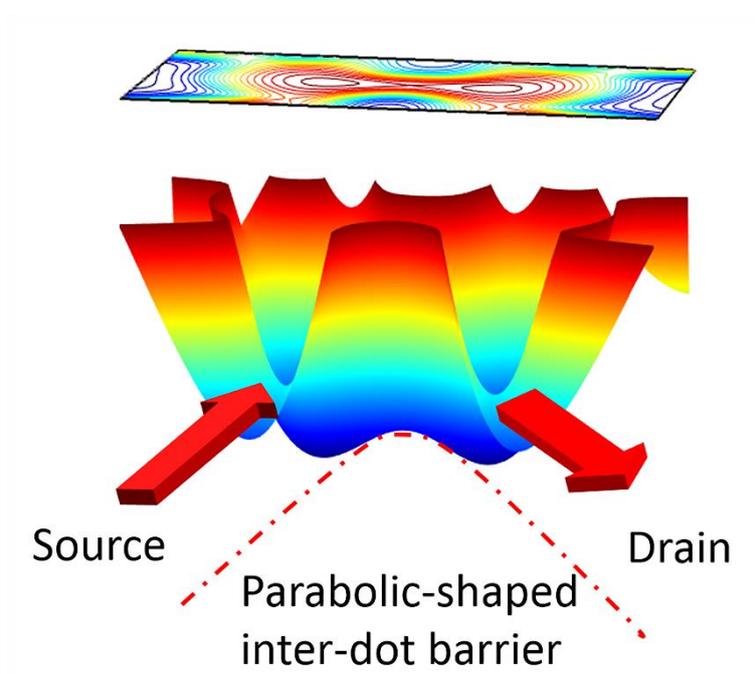

**fig. S4.** COMSOL simulation of the inter-dot barrier.

## 4. COMSOL simulation for different values of $V_{DM}$

When tuning $V_{DM}$ to more negative values, the inter-dot potential barrier arises, leading to the reduction of the coupling strength between the dots (as shown in Fig. 3). A COMSOL simulation is used to calculate the change of potentials for different values of $V_{DM}$, while other gate voltages remain fixed, as shown in fig. S5. The inter-dot barrier increases when tuning $V_{DM}$ more negative. Meanwhile, the dot confinement potential remains almost unaffected.

The schematic diagram of the evolution of such double-dot to single-dot transition of the confining potential at the crossline in fig. S5 is shown in Fig. 3, C to E. Such simulation results agree well with the experiment results.

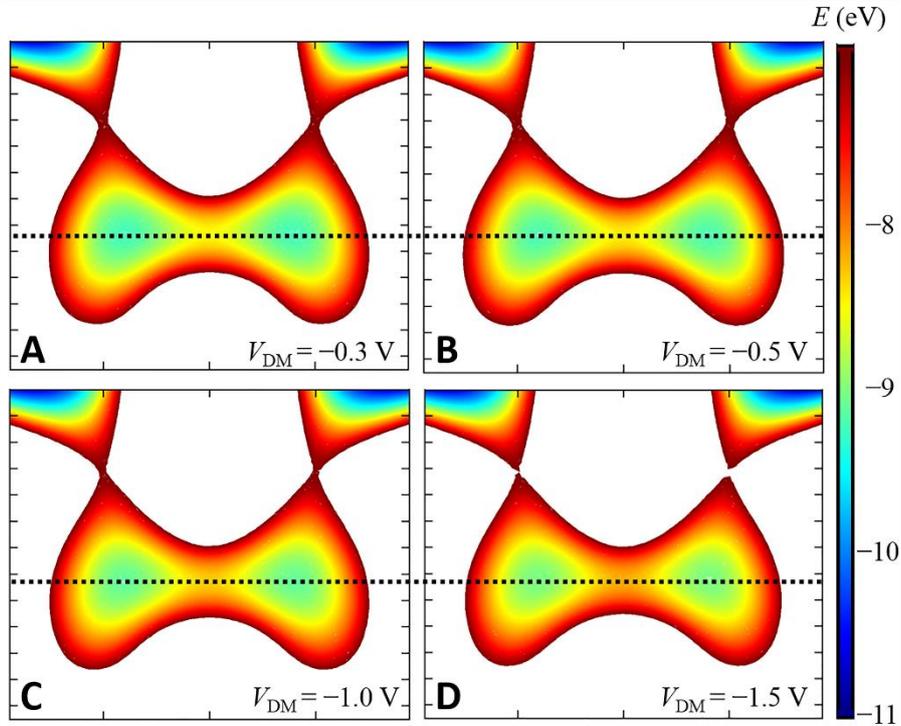

**fig. S5.** COMSOL simulation on the potential-well distribution of the closed contours shown in Fig. 2D based on the designed pattern for $V_{BG}$ = 30 V, $V_{LP}$ = $V_{RP}$ = 0 V, $V_{LB}$ = $V_{RB}$ = −1.5 V, $V_{UM}$ = −2.1 V, and $V_{DM}$ = −0.3 V, −0.5 V, −1 V and −1.5 V for **A** to **D**, respectively.

## 5. Gate controllability in the low-density regime

Because of the different values of $V_{BG}$, the formation of the double quantum dot was dominated by different mechanisms, as shown in Fig. 4, D and E. At a relatively low Fermi energy ($E_{F1}$), the intrinsic and fabrication-induced impurities dominate the confining potentials of the transport behavior, which cannot be well controlled by electrostatic gating. The controllability of the electrostatic gates here is demonstrated in fig. S6, A to C. When tuning the value of $V_{DM}$ and $V_{UM}$ together over a wide range, the tunneling rate between the source/drain and the dot changes effectively, while the coupling strength of two quantum dots does not show any obvious signature of evolvement.

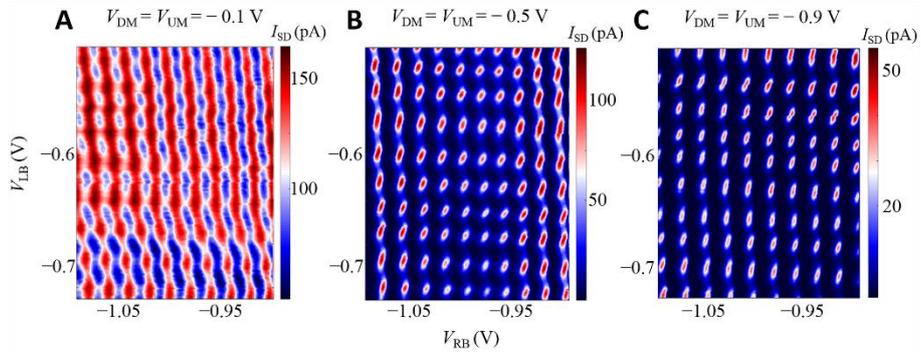

**fig. S6.** Current through the double quantum dot versus $V_{LB}$ and $V_{RB}$ applied to the gates LB and RB for $V_{BG}$ = 25 V, $V_{LP}$ = $V_{RP}$ = 0 V, bias voltage at $V_{SD}$ = 100 μV and $V_{DM}$ = $V_{UM}$ = −0.1 V, −0.5 V, and −0.9 V for **A** to **C**, respectively.